\documentclass[aps,prb,twocolumn,groupedaddress,showpacs]{revtex4-1}
\usepackage{graphicx}
\usepackage{dcolumn}
\usepackage{bm}
\usepackage{color}
\usepackage{url}
\usepackage[colorlinks=true,urlcolor=blue,linkcolor=blue,citecolor=blue]{hyperref}

\begin{document}

\title{Shape and Energy Consistent Pseudopotentials for Correlated Electron systems}

\author{J. R. Trail}\email{jrt32@cam.ac.uk}
\affiliation{Theory of Condensed Matter Group, Cavendish Laboratory,
  J J Thomson Avenue, Cambridge CB3 0HE, United Kingdom}

\author{R. J. Needs}
\affiliation{Theory of Condensed Matter Group, Cavendish Laboratory,
  J J Thomson Avenue, Cambridge CB3 0HE, United Kingdom}

\date{\today}

\begin{abstract}

 A method is developed for generating pseudopotentials for use in
 correlated-electron calculations.
 The paradigms of shape and energy consistency are combined and defined
 in terms of correlated-electron wave-functions.
 The resulting energy consistent correlated electron pseudopotentials (eCEPPs)
 are constructed for H, Li--F, Sc--Fe, and Cu.
 Their accuracy is quantified by comparing the relaxed molecular geometries and
 dissociation energies they provide with all electron results, with all
 quantities evaluated using coupled cluster singles doubles and triples
 calculations.
 Errors inherent in the pseudopotentials are also compared with those arising
 from a number of approximations commonly used with pseudopotentials.
 The eCEPPs provide a significant improvement in optimised geometries and
 dissociation energies for small molecules, with errors for the latter being an
 order-of-magnitude smaller than for Hartree-Fock-based pseudopotentials available
 in the literature.
 Gaussian basis sets are optimised for use with these pseudopotentials. 

\end{abstract}

\pacs{71.15.Dx, 02.70.Ss, 31.15.V-}


\maketitle

\section{Introduction \label{sec:introduction}}

Electronic structure methods in solid state theory and quantum chemistry
provide a powerful hierarchy of \emph{ab initio} tools for the accurate description 
of interacting atomic systems.
There are fundamental limits to the accuracy at each level of theory.
Pseudopotentials, or effective core potentials, provide an approximation that
replaces the interaction of largely inactive core electrons and valence
electrons with a potential, which reduces the number of particles 
required.\cite{Dolg_2012}
Such an approximation is often necessary due to limitations on computational 
resources, and it also allows the elimination of analytic and numerical problems 
associated with singularities in electron-nuclear interaction potentials.

Methods for generating pseudopotentials within density functional theory
(DFT) are well established.
Errors associated with pseudopotentials are often systematically controlled,
and reliably estimated to be small and secondary to the errors due to 
approximate exchange correlation functionals.\cite{Milman_2000}
However, the accuracy of DFT is not sufficient for many systems.
Methods that involve explicitly correlated wave functions,
such as coupled cluster\cite{ccsd} (CC) or quantum Monte
Carlo\cite{Ceperley-Alder_1980,Foulkes_QMC_review,Casino_reference,
Lester_review_2012} (QMC) theory, are often required.

The cost of coupled-cluster calculations with single, double, and
perturbative triple excitations [CCSD(T)] scales with the seventh
power of the number of electrons, or as $O(n^3N^4)$,\cite{ccsd} with
$n$ the number of electrons and $N$ the number of basis functions.
The use of pseudopotentials reduces both $n$ and $N$ by removing core
electrons and requiring fewer basis functions than the
nucleus-electron interaction.

The computational cost of fermion QMC calculations increases more
slowly, as the third or fourth power of the number of particles.
However, the scaling of the error with atomic number, $Z$, is
$Z^{5}-Z^{6.5}$.\cite{Ceperley_1986,Ma_2005} The use of
pseudopotentials reduces the effective value of $Z$, making QMC
calculations feasible for systems with heavy atoms.

At these higher levels of theory estimates of, and methods to control,
pseudopotential errors are not usually available.
Furthermore, available pseudopotentials often reproduce atomic
properties at the DFT or HF level of theory.
This introduces an uncontrolled error due to the inconsistent application of
theory to the definition and application of the pseudopotentials.

In this study we generate pseudopotentials from explicitly correlated
atomic wave-functions, with no recourse to reducing these many-body
wave-functions to orbitals arising from a self-consistent potential.
The influence of core electrons on valence electrons is recreated by
consistently combining three distinct representations of the core-valence
interaction, namely core scattering, core polarisation, and differences between
ground state energies.

Scattering properties of core electrons are defined in terms of density 
matrices of atoms or ions with one valence electron.
Long-range effects from dynamic core relaxation (arising from
electron-electron correlation) can be included in the pseudopotential via
a core polarisation potential (CPP).\cite{Mitroy_2010}

Pseudopotentials defined to reproduce the one-body core-valence
interaction and core polarisation are not unique.
Consequently, the accuracy may be further improved by searching over all
pseudopotentials that reproduce such properties, and seeking that which most
accurately reproduces ionisation and excitation energies, and electron
affinities for an isolated atom.

A combined reproduction of core scattering, core polarisation, 
and atomic excitation energies allows the generation of a new
pseudopotential from correlated electron calculations,
referred to as an energy consistent correlated electron
pseudopotential (eCEPP).

Our procedure for constructing eCEPPs is an extension of the correlated electron
pseudopotential (CEPP) method described in two previous papers,
\cite{Trail_2013_pseudopotentials,Trail_2015_pseudopotentials}
which are summarised here.
Only two of the criteria given above are used to generate CEPPs,
the preservation of core scattering and the representation
of core polarizability.
An all-electron (AE) multi-determinant wave-function for an ion/atom with one
valence electron provides an AE charge density.
From this a single-electron charge density is constructed by removing an
\emph{ab initio} core electron charge density.
Within a `core radius' of the nuclei the single-electron charge density 
is replaced by a standard form that preserves the continuity of the value
and first four derivatives at the core radius.
Direct inversion of a one-electron Schr\"odinger equation then provides
the pseudopotential.
The analytic basis for this approach is an extension of
norm-conservation\cite{Hamann_pseudopotentials_1979} to the many-body
Hamiltonian, which conserves the scattering properties of core electrons to
first-order by reproducing the one-body density matrix outside of the core
region.\cite{Acioli_1994}

Two subtle issues arise in the process of constructing CEPPs.
The multi-determinant wave-function is composed of orbitals that are not
unique, and without associated energy eigenvalues.
A separation of the density matrix into a core and valence components
is achieved using natural orbitals,\cite{Davidson_1972} and defining
the core in terms of the natural orbitals with the largest occupation numbers.
While core natural orbitals have no direct effect on the pseudo
wave-function in the core region they validate the core radius used and remove
a small but significant influence of core electrons at the core radius.

The long-range, one-body part of the CPP naturally emerges in the process 
of generating a CEPP.
This is represented using well-known parameterised CPP
forms\cite{Shirley_1993,Muller_1984} such that the
CEPP is the sum of an effective potential and the CPP, and the 
one-body part of the total potential is \emph{ab initio}.
The ion-ion and many-body part of the CPP remains semi-empirical, but 
with parameters that are consistent with the one-body component.

The CEPPs generated for H, Li--F, and Sc--Fe, together with CCSD(T), were 
shown to reproduce AE results for small molecules.
Perhaps the strongest indicator that the CEPP provides a good starting point is
that it is a pseudopotential generated from a highly ionic state which is
accurate for a neutral molecule.
This is in contrast to the well known poor transferability of KS-DFT
pseudopotentials between different oxidation states, which is primarily caused
by a pseudopotential definition that takes the self-consistent potential to be
fixed and uses de-screening to remove the interaction between valence
electrons.
Neither of these approximations is used for the CEPPs.
Iron provides a good example, with a CEPP generated from a Fe$^{15+}$ ion
resulting in properties for molecules containing iron which are as accurate as
for molecules with few valence electrons such as Li$_2$.

Before defining eCEPPs we note two alternative strategies for
improving the accuracy of pseudopotentials that have not been investigated.
The definition of the CEPPs could be extended to reproduce 
two-body core scattering properties.
This would require the construction of a pseudo-Hamiltonian that 
reproduces both first and second-order density matrices outside the core region
of an isolated ion/atom with two valence electrons.
Such a pseudo-Hamiltonian would contain an extra two-body potential 
that describes the electron-electron interaction mediated by core electrons.
This two-body interaction potential is approximately included in the CPP
part of the CEPPs and is small, so it is unlikely that a more accurate
description would significantly improve the pseudopotential accuracy.

Another, related, strategy would be to reproduce one-body core scattering
properties to higher than first-order.
An extension of norm-conserving KS-DFT pseudopotentials to reproduce higher
order scattering properties\cite{Shirley_1989} provides a modest reduction in
transfer error, suggesting that this approach will not reliably provide
a more transferable pseudopotential.

These observations suggest that an eCEPP that conserves the CEPP properties
while enforcing further `energy consistency' between a number of states may
provide more accurate pseudopotentials.
Such a combination of extrapolation and interpolation of atomic properties 
combines the strengths of shape-consistent and energy consistent 
\emph{ab initio} pseudopotentials.
Electron correlation can be included throughout, without the need for
an independent electron approximation.

The article is organised as follows.
Section\ \ref{sec:eCEPPs_theory} describes the basis and implementation of the
combined energy and shape consistent strategy outlined above.
The ability of eCEPPs to reproduce AE results in correlated electron
calculations is analysed in Sec.\ \ref{sec:results}.
The performance of the pseudopotentials for titanium are addressed in Sec.\
\ref{sec:titanium_atom}, and errors in optimised geometries and dissociation
energies for a test set of molecules composed of first row and $3d$ transition
metal atoms are assessed in Sec.\ \ref{sec:molecules}.
Section\ \ref{sec:errors} provides a comparison of errors arising from 
the eCEPPs with those from other approximations commonly used in
\emph{ab initio} calculations.
Gaussian basis sets optimised for use with these potentials are described 
and validated in Sec.\ \ref{sec:eCEPP_basis}.

\section{Shape and energy consistent correlated pseudopotentials \label{sec:eCEPPs_theory}}

\subsection{Theoretical basis \label{sec:method}}

The pseudopotentials are formulated as the sum of a semi-local operator and a
many-body potential,
\begin{eqnarray}
\label{semiloc}
\hat{V}^{\text{pp}} &=& \sum_0^{l_{max}-1} \sum_{m=-l}^{l}
     | Y_{lm} \rangle \left[ V^{\text{pp-cpp}}_l - V^{\text{pp-cpp}}_{l_{max}} \right] \langle Y_{lm} | \nonumber \\
 & & + V^{\text{pp-cpp}}_{l_{max}} + V^{\text{cpp}},
\end{eqnarray}
where each $V^{\text{pp-cpp}}_l$ and the $V^{\text{pp-cpp}}_{l_{max}}$ act
on the spherical harmonic projection to channels $l<l_{max}$ and $l \ge l_{max}$,
respectively.
The potential $V^{\text{cpp}}$ is the many-body CPP.

The $V^{\text{pp-cpp}}_l$ terms are represented using the parameterised form
\begin{eqnarray}
\label{pp_param1}
\tilde{V}^{\text{pp-cpp}}_l &=& \sum_{q=1}^{6} A_{ql} r^{n_{ql}} e^{-a_{ql}r^2} \nonumber \\
                            &=&
\left\{ \begin{array}{rl}
 V^{\text{pp-cpp}}_l - V^{\text{pp-cpp}}_{l_{max}}    & l \neq l_{max} \\
 Z_v/r               + V^{\text{pp-cpp}}_{l_{max}}    & l =    l_{max} .
\end{array} \right.
\end{eqnarray}
For the local channel ($l=l_{max}$), $n_{ql}=-1$ for $q=1$, $n_{ql}=+1$ for
$q=6$, and $n_{ql}=0$ otherwise.
For other channels $n_{ql}=0$.
The pseudo-atomic number is $Z_v=Z-Z_c$, where $Z_c$ is the number of core
electrons represented by the pseudopotential.
The local channel is $l_{max}=max\{l_{occ}\}$, 
where $\{l_{occ}\}$ is the set of all $l$ values occurring in
the electronic configurations used to generate the pseudopotential.

In terms of the parameterised components of the pseudopotential, 
$\tilde{V}^{\text{pp-cpp}}_l$, the complete pseudopotential is
\begin{equation}
\label{pp+cpp}
V^{\text{pp}}_{l} = 
\left\{ \begin{array}{rl}
\tilde{V}^{\text{pp-cpp}}_{l} -Z_v/r + \tilde{V}^{\text{pp-cpp}}_{l_{max}} + V^{\text{cpp}} & l \neq l_{max} \\
-Z_v/r  + \tilde{V}^{\text{pp-cpp}}_{l_{max}}                              + V^{\text{cpp}} & l =    l_{max} .
\end{array} \right.
\end{equation}
The $V^{\text{cpp}}$ potential is composed of the sum of 
a one-body core-electron potential, a two-body electron-electron
interaction, and two further terms involving the positions of both electrons
and nuclei which are expressed in the form provided by Shirley and
Martin.{\cite{Shirley_1993}
For a single atom with one valence electron only the local component of
the CPP remains,
\begin{equation}
\label{cpp_local}
V^{\text{cpp}}= -\frac{1}{2} f^2(r/r_0) \frac{\alpha}{r^4}.
\end{equation}
The short range truncation function $f(x)=(1-e^{-x^2})^2$ removes the
non-physical singularity at $r=0$, $r_0$ is a cutoff radii for this function,
and $\alpha$ is the dipole polarizability of the core.

The previously available CEPPs are parameterised in the same
form.
An eCEPP candidate is expressed in terms of scaled CEPP parameters,
\begin{eqnarray}
A_{ql} &=& \lambda^A_{l}  A^{CEPP}_{ql}  \\
a_{ql} &=& \lambda^a_{ql} a^{CEPP}_{ql}
\end{eqnarray}
and $r_0$ is varied without scaling, so that the candidate potentials are equal
to the CEPP for scaling parameters $( \{\lambda^A_{l}\} , \{\lambda^a_{ql}\} )$
all equal to unity and $r_0=r_0^{CEPP}$.

Overcompleteness, and the accompanying emergence of undesirable oscillations
in the potential, is prevented by the uniform scaling of $\{A_{ql}^{CEPP}\}$
parameters.
The CPP cutoff radii, $r_0$, is taken to be a free parameter,
while the core polarizability, $\alpha$ in Eq.\ (\ref{cpp_local}), 
is taken from previously published highly accurate values available for all
the atoms considered.\cite{Shirley_1993,Patil_1985}

The pseudopotential is required to be finite and have a first- and second
derivative equal to zero at $r=0$.
Taking these constraints into account results in
$8 + 7l_{max}$ free parameters (for first row and $3d$ transition metal
atoms $l_{max}=$2 and 3, resulting in 22 and 29 parameters,
respectively).

In order to seek an eCEPP we start with all scaling parameters
equal to unity.
Scaling parameters are then varied to search for new values
that preserve the CEPP charge density outside of the core region for
single-valence electron states, and reproduce AE energy differences between
atomic states of different symmetry and charge.
The relaxation of the one-electron pseudo charge density within the core region
provides the freedom for a candidate potential to reproduce the AE energy
differences while preserving the core-valence interaction that defines the CEPP.

We define a penalty function that is zero for an eCEPP, and expressed
as a sum of two terms, $P=P_1+P_2$.
The first of these, $P_1$, measures the degree
to which a candidate potential reproduces the one-electron density
outside of the core region.  This is given by
\begin{widetext}
\begin{equation}
\label{cost_func1}
P_1(\{\lambda^A_{l}\},\{\lambda^a_{ql}\},r_0) = \sum_{l=0}^{l_{max}} \left(
     \left[ \phi_l(r^c_l) -  \rho^{\frac{1}{2}}_l(r^c_l) \right]^2 + 1 -
     \left[ \int_{r^c_l}^{\infty} \phi_l \rho^{\frac{1}{2}}_l dr \right] ^2 \Big/
     \left[ \int_{r^c_l}^{\infty} \phi_l^2 dr
            \int_{r^c_l}^{\infty} \rho_l dr           \right]
\right)
\end{equation}
\end{widetext}
where the sum is taken over the single-valence-electron $^2L$ states
corresponding to $l=L$ for $l \leq l_{max}$, and the core
radius for each channel, $r^c_l$, defines the sphere within which the
one-electron pseudo charge density is allowed to vary.

In Eq.\ \ref{cost_func1}, $\rho_l$ is the AE electron density with the
core electrons removed, and $\phi_l$ is the one-body wave-function
corresponding to the candidate pseudopotential.  For the initial
parameter values, $\rho_l=\phi^2_l$ and $P_1=0$.

Equation\ (\ref{cost_func1}) is constructed as the simplest form for
which $P_1$ is non-negative, the lowest order variation of $P_1$ about
the minimum is second order, and $\rho_l=\phi^2_l$ outside of the core
region results in $P_1=0$.

A minimum with $P_1=0$ also occurs for parameter values that are not equal
to those for the CEPP.
It is these parameters that we seek, as they correspond to the many 
pseudopotentials for which the one-electron charge densities are equal to the
AE valence charge density outside of the core region only.

The second component of the penalty function, $P_2$, measures the degree to
which the candidate potential reproduces atomic excitation energies.
A set of $N+1$ states is considered, which is generally composed of the neutral
atom ground state, the anion ground state, and several ground and excited
states for neutral atoms and ions.
We index the set of states using $j$, with $j=0$ the neutral ground state, and
define the second penalty function as

\begin{widetext}
\begin{equation}
\label{cost_func2}
P_2(\{\lambda^A_{l}\},\{\lambda^a_{ql}\},r_0) = \frac{1}{\beta^2} \times
     \sum_{j=1}^{N}
     \left[
     \left( E_j^{\text{AE}} - E_0^{\text{AE}} \right) - \left( E_j^{\text{pp}} - E_0^{\text{pp}} \right)
     \right]^2 
\end{equation}
\end{widetext}
where $E_j^{\text{AE}}$ and $E_j^{\text{pp}}$ are the total energies of
the $j^{th}$ atomic state evaluated with all electrons present and the
candidate pseudopotential, respectively.
The prefactor $1/\beta^2$ sets the importance of the second penalty
function relative to the first, with $\beta$ setting the accuracy with
which the eCEPP is required to reproduce AE total energy differences.
(No prefactor is used in $P_1$ as it is expected to be bounded by 0 and
${\sim}1$ for all parameter values.)

Minimising the full penalty function,
$P(\{\lambda^A_{l}\},\{\lambda^a_{ql}\},r_0)$,
with respect to $( \{ \lambda^A_{l} \}, \{ \lambda^a_{ql} \}, r_0 )$
provides the eCEPP that preserves core scattering properties 
as well as reproducing the AE energy spectrum of an isolated atom.

\subsection{Implementation \label{sec:implementation}}

The evaluation of the two components of the penalty function, $P=P_1+P_2$, are 
distinct from each other.

The function $P_1$ includes correlation effects if the target
charge density, $\rho_l$, is provided by a correlated-electron
calculation.
We take the target density outside of the core region to be that provided by
the CEPP as, by definition, this is equal to the AE charge density provided by
Multi-Configuration Hartree Fock (MCHF)\cite{Fischer_1997,atsp2k} calculations
with the core contribution removed.
Consequently, the target and candidate charge densities outside of the core
are both obtained by direct solution of the one-body Schr\"odinger equation.
The target charge density is provided using the CEPP, while
the candidate charge density is provided by the candidate pseudopotential.
Each term in $P_1$ is evaluated using summation and numerical integration.

The evaluation of $P_2$ is performed by summation of the squared differences 
between CCSD(T) energies, with energies evaluated using Gaussian basis sets
and extrapolation to the complete basis set (CBS) limit.

All CCSD(T) energies are evaluated using the \textsc{MOLPRO}\cite{molpro} code
for both AE and candidate eCEPP atoms to provide the target and candidate
energies, respectively.
The active space is defined to include all electron excitations, including
core electrons when they are present.
Uncontracted (relativistic) aug-cc-pV$n$Z(-DK)\cite{basis,basisDK} basis sets are
used to evaluate (AE) energies, where $n$ is the number of correlating functions
present in the basis.
Differences between total energies, such as electron affinities and ionisation
energies, are provided in the CBS limit using extrapolation
\cite{Trail_2013_pseudopotentials,Trail_2015_pseudopotentials,Feller_2010}
and CPP corrections when required.
Extrapolating energies from $n=Q,5$ basis sets and correcting to include CPP
energies from $n=T$ basis sets is referred to as `$n=(TQ5)$
extrapolation'.\cite{Supplemental}

\subsection{eCEPPs for H, Li--F, Sc--Fe, and Cu\label{sec:eCEPP_atoms}}

Energy-consistent correlated electron pseudopotentials are constructed for the
atoms H, Li--F, Sc--Fe and Cu.
This is the set of atoms for which CEPPs are available, with Cu added due to
its many properties and applications.
There is no core for H, for Li--F a [He] core is represented by the
pseudopotentials, and for Sc--Fe and Cu the core is [Ne].
The initial CEPP parameter values are taken from previous
publications, except for Cu for which a CEPP is generated using the same method.
Core radii for each pseudopotential channel are the same as for the
CEPPs.\cite{Trail_2013_pseudopotentials,Trail_2015_pseudopotentials}
The total of core natural orbital occupation numbers arising when generating 
CEPPs are close to the total number of core electrons, deviating by $7$--$1
\times 10^{-3}$ and $6$--$2 \times 10^{-3}$ for the first row and $3d$
transition metal atoms, respectively.
Across each row this small deviation decreases monotonically with $Z$, and is
correlated with the CEPP (and eCEPP) core radii.

For first row atoms $P_1$ is defined using the $^2S$, $^2P$, $^2D$ 
single-valence ground-states, while for the $3d$ transition metal atoms
the $^2F$ state is also included.

A finite number of states are used to define $P_2$.
Selection rules provide $N+1=$5--8 states for the first row atoms, and 
$N+1=$8--9 states for the $3d$ transition metal atoms.\cite{Supplemental}

Our target accuracy for the eCEPPs is better than chemical accuracy, so we
take the parameter $\beta$ in Eq.\ (\ref{cost_func2}) to be 43 meV.
The combination of parameters, numerical integration for $P_1$, and
CCSD(T) data extrapolated to the CBS limit for $P_2$ provides the penalty
function for a given set of free parameters, $P( \{\lambda^A_{l}\} ,
\{\lambda^a_{ql}\} , r_0)$.

Optimisation of the penalty function is carried out using the
Broyden$-$Fletcher$-$Goldfarb$-$Shanno (BFGS) method,\cite{Kelley_1999}
with the transformed parameters
 $( \{ \lambda^A_{l} \} , \{ \ln \lambda^a_{ql} \} , r_0)$
used to improve stability.
Numerical derivatives are defined with a finite difference of
$\Delta=10^{-5}$, the smallest value possible given the precision
of the CCSD(T) calculations.

Energies and derivatives for the first row atoms are evaluated using
the $n=(TQ5)$ extrapolation.  However, for the $3d$ transition metal
atoms a slightly different approach was required.  Convergence with
basis set is slow and extrapolation using $n=(TQ5)$ is required to
achieve the target accuracy.  However, although this choice of basis
sets provides CCSD(T) energies converged to the required precision, a
significant numerical error in the finite difference prevents the
optimisation method from succeeding reliably. 

Reliable convergence to the required accuracy is achieved by
evaluating the penalty function using $n=(TQ5)$ extrapolation, but
evaluating the \emph{gradient} of the penalty function using energies
provided by the $n=Q$ basis only 
(for both AE and candidate eCEPP total energies).
This stabilises and accelerates the optimisation process.

The optimisation was terminated for $P<N/400$.
This criterion provides a mean absolute agreement between eCEPP and AE 
energy differences of 3 meV, with a maximum disagreement 
of 11 meV occurring for ionisation to the $^6S$ state of
Cr$^{+}$.
This optimisation error is small when compared with the estimated CCSD(T)
basis set error for all atoms.

Optimisation provides an eCEPP for which the penalty function 
is small, but non-zero.
This corresponds to the charge density of the single valence electron ion 
relaxing outside of the core region and not exactly reproducing 
the target MCHF AE density.
The degree of relaxation is small and adequately controlled by the
penalty function chosen, with the overlap term differing from unity by
$<2 \times 10^{-4}$ for all atoms and states (the maximum occurs for
the $^2D$ state of Sc).  The mean-absolute error for $\phi(r^c_l)$ is
$0.6\%$, with all errors less than $3.5\%$.

Pseudopotentials for H are unusual in that the CEPP is equivalent to a
norm-conserving HF pseudopotential since neither correlation or exchange occur
and no CPP arises.
Furthermore, only a few bound atomic states are available and the CEPP
accurately reproduces the energies of these states.
Consequently, for H the CEPP and eCEPP are identical and equal to a
norm-conserving HF pseudopotential.

\section{Results \label{sec:results}}

\subsection{Titanium atom and Titanium Oxide molecule \label{sec:titanium_atom}}

\begin{figure}[t]
\includegraphics[scale=1.00]{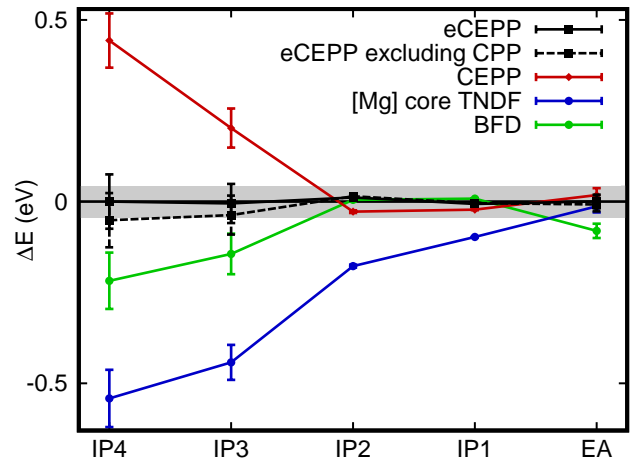}
\caption{ \label{fig1} 
The difference between atomic Ti ionisation and electron affinities 
evaluated using pseudopotentials, and those from relativistic AE calculations.
The ionisation energies IP4 to IP1 and the electron affinity EA 
are energy differences between ground state total energies 
for a total charge of $+4$ to $-1$.
The grey region indicates chemical accuracy 
of 1 kcal$.$mol$^{-1}$ = 43 meV, and the error bars 
are for the estimated CBS limit.
All energies are evaluated using CCSD(T).
}
\end{figure}

\begin{figure}[t]
\includegraphics[scale=1.00]{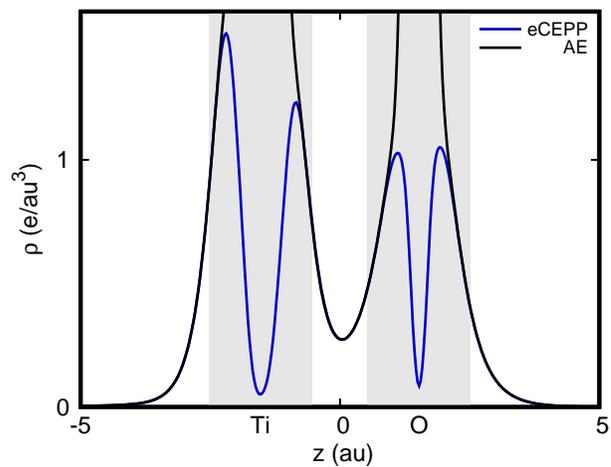}
\caption{ \label{fig2} 
The charge densities for TiO along the axis containing both nuclear sites.
Densities are provided by AE and eCEPP, with CISD.
The grey areas contain the core regions for each atom.
}
\end{figure}

Two basic requirements of an accurate pseudopotential are to reproduce AE
energy differences between atomic states and to reproduce charge densities
outside of the core region.

Figure~\ref{fig1} shows the deviation of pseudo-atomic energies from AE
energies for several Ti pseudopotentials (including the eCEPP).
Energy differences shown are ionisation energies and the electron affinity,
evaluated as differences between CCSD(T) energies extrapolated to
the CBS limit.

The eCEPP energies agree with AE results 
to better than both CBS error and chemical accuracy.
This is expected in light of the generation procedure.
Invoking the eCEPP with no CPP included introduces a small error
due to the low polarizability of the [Ne] core.

The figure also provides a useful comparison of the accuracy of the
eCEPP with the
Trail-Needs Dirac-Fock\cite{Trail_2005_pseudopotentials,TNDF_website} (TNDF)
norm-conserving pseudopotential, the
Burkatzki-Filippi-Dolg\cite{Burkatzki_2007,BFD_website} (BFD) energy consistent
HF pseudopotential, and the CEPP.
For all four pseudopotentials the CBS error consistently increases with
ionisation, as expected given that the aug-cc-pV$n$Z(-DK) basis sets
are optimised for neutral atoms.
Although the CBS error is largely removed by de-contracting basis sets, 
the width parameters of the Gaussian basis become increasingly sub-optimal
as charge densities contract with increasing total charge.

All three non-eCEPPs show an increase in pseudopotential error with
increasing charge.
This trend is caused by the cancellation of errors in energy
differences becoming less perfect with increasing energy difference,
and is particularly apparent for the large ionisation energies
where electrons are removed from the $4s$ sub-shell, (IP3 and IP4).

Comparing errors arising for all pseudopotential types suggests
an interpretation of the errors removed from the eCEPPs.
The large errors for the TNDF pseudopotentials suggests that a [Ne] core is
necessary for $3d$ transition metal atoms.\cite{Pacios_1988}
The large errors for the BFD pseudopotential suggest that energy-consistency at
the CCSD(T) level of theory is required, and HF is not sufficient.
The opposite sign, and similar magnitude, of the errors for the 
CEPP and BFD pseudopotentials suggests that a combination of 
CEPP and energy consistency is likely to be successful.
The accuracy provided by the eCEPP confirms that it successfully removes these
errors.

The eCEPP is not constrained to reproduce the charge density for any system
except the single-valence ion, so we compare the AE and pseudo charge densities
for a small, neutral molecule.
Figure\ \ref{fig2} shows such charge densities for the TiO molecule, with an
experimental bond length.\cite{Harrison_2000}
Densities are evaluated using configuration interaction singles and
doubles (CISD) calculations, and plotted along the primary
symmetry axis.
The densities provided by the AE and eCEPP calculations agree well outside of
the core region for each atom, demonstrating that good transferability is not
limited to atomic ionisation energies.

These results confirm that the eCEPP for titanium accurately reproduces the
atomic properties it is designed to recreate, and that the pseudopotentials
transfer well to the TiO molecule, accurately reproducing AE charge densities
outside of the core regions.

\subsection{Small molecules geometries and dissociation energies \label{sec:molecules}}

\begin{figure*}[t]
\includegraphics[scale=1.00]{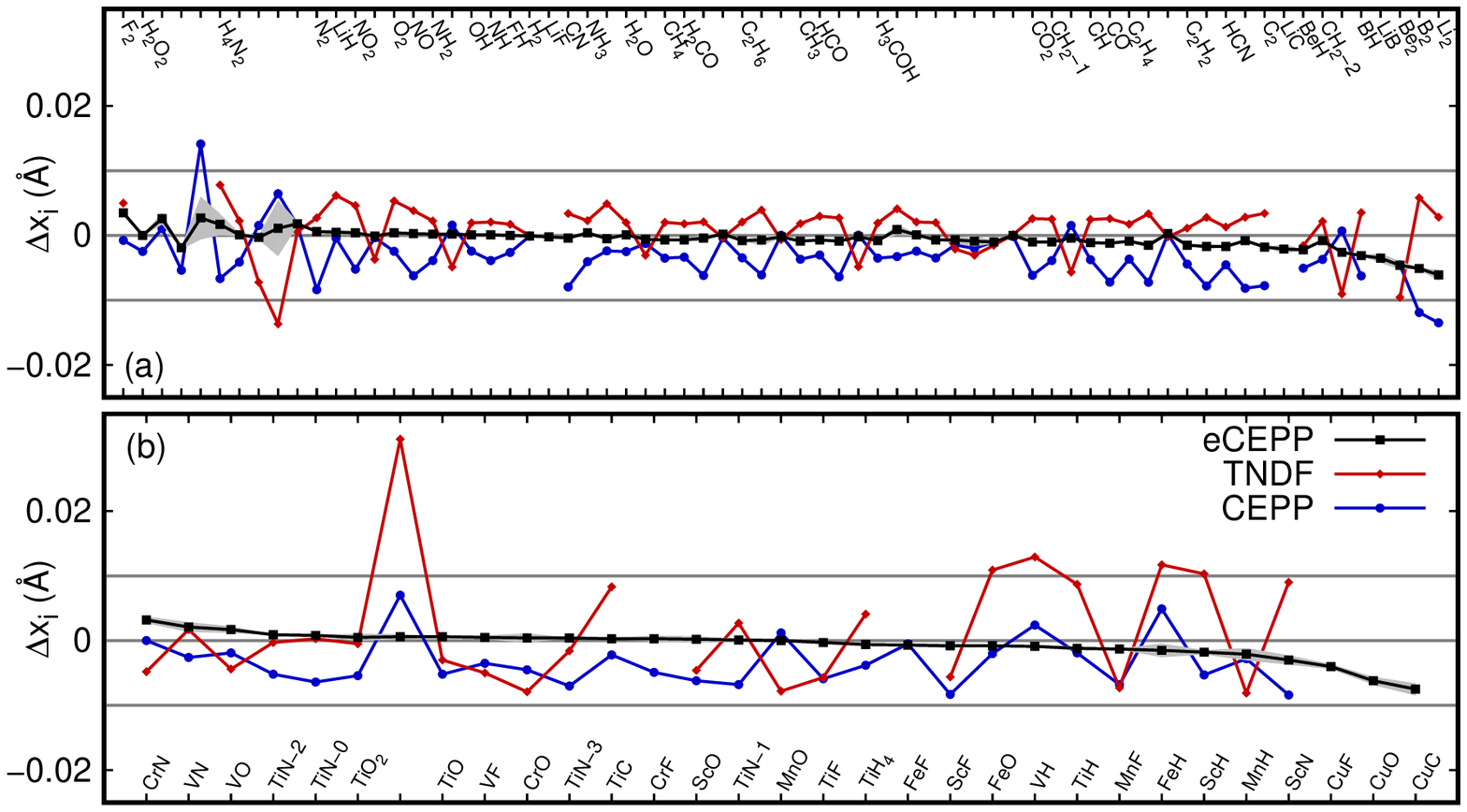}
\caption{ \label{fig3} 
The difference between optimised molecular geometry parameters, $x_i$, 
from pseudopotentials and AE CCSD(T) calculations.
The top bar shows differences for molecules containing 1$^{st}$ row atoms only,
and the bottom bar shows differences for molecules containing $3d$ transition
metal atoms.
Data for the CEPP and TNDF pseudopotentials are from
[\onlinecite{Trail_2013_pseudopotentials}] and
[\onlinecite{Trail_2015_pseudopotentials}].
Bond and dihedral angle parameters are expressed as the arc-length on a circle
of radius $1.0$\AA.
Grey lines at $\pm$0.01\AA\ contain all data points that reproduce
AE geometries to within chemical accuracy.
Labels absent from the horizontal axes mark further geometry parameters of the
molecule labelled immediately to the left.
}
\end{figure*}

\begin{figure*}[t]
\includegraphics[scale=1.00]{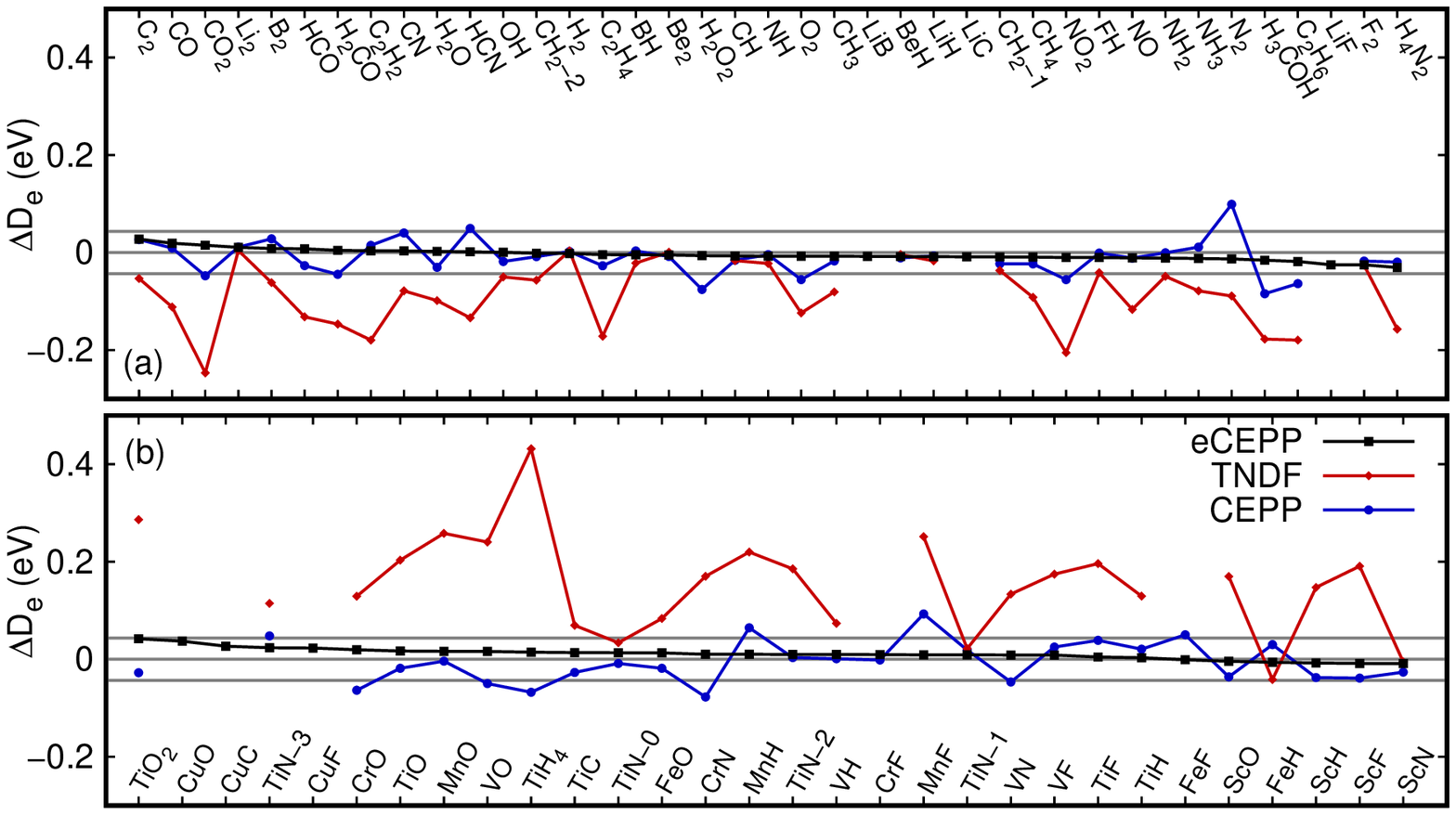}
\caption{ \label{fig4} 
The difference between molecular dissociation energies, $D_e$, from
pseudopotential and AE CCSD(T) calculations.
The top bar shows differences for molecules containing 1$^{st}$ row atoms only,
and the bottom bar shows differences for molecules containing $3d$ transition
metal atoms.
Data for the CEPP and TNDF pseudopotentials are from
[\onlinecite{Trail_2013_pseudopotentials}] and
[\onlinecite{Trail_2015_pseudopotentials}].
The grey lines at $\pm$43 meV contain all data points that
reproduce AE energies to within chemical accuracy.
}
\end{figure*}

To assess the accuracy of the eCEPPs in reproducing molecular
dissociation energies and equilibrium geometries we consider a
moderately large set of small molecules.
A test set is constructed starting with the neutral members of the
G1 set\cite{G1} which contain only the atoms H and Li--F.
We then add LiB, LiC, LiF, H$_2$, BH, Be$_2$, B$_2$, C$_2$, and NO$_2$,
to give 38 first row molecules (in 39 states).
To this set we add the diatomic molecules composed of Sc--Fe and Cu
together with H, C, N, O, and F atoms, selecting those for which the
ground state term and dominant configuration is not in doubt, as
described by Harrison,\cite{Harrison_2000} and for which
single-reference CCSD(T) is stable.
Titanium is of particular interest,\cite{Titanium_tech,Trail_2017} so we also
include TiH$_4$, TiO$_2$, and three excited states of TiN.
The final test set contains 65 molecules in 69 states.\cite{Supplemental}

Geometry optimisation is performed for each molecule by minimisation of
the AE and pseudopotential CCSD(T) energies with respect to 
bond lengths and bond--angles that characterise the geometry
of each molecule, $x_i$.
All $x_i$ are corrected to include the influences of the CPP,
and the basis set error estimated.\cite{Supplemental}

The differences between molecular geometry spatial parameters 
arising from the eCEPP and AE Hamiltonians are shown in
Fig.\ \ref{fig3}.
The figure also shows the geometry parameters previously reported
for the CEPP and TNDF pseudopotentials when available.

All eCEPP geometry parameters fall within chemical accuracy (of 0.01\AA) of
the relativistic AE values, unlike the results for both the TNDF and CEPP
pseudopotentials.

The eCEPP geometry parameters for the mean absolute deviation (MAD)
from AE results is ${\sim}1/4\times$ that for either the CEPPs or TNDF
pseudopotentials.  The overall bias is small for the eCEPPs, with an
average deviation of $6\times10^{-4}$\AA.

For all molecules considered the estimated basis set error is
negligible, except for H$_2$O$_2$ and H$_4$N$_2$ for which a
particularly shallow Born-Oppenheimer surface magnifies this error for
certain angular degrees of freedom.  There is a weak correlation
between errors for the TNDF pseudopotentials and CEPPs, but no
correlation remains in the eCEPP errors.

For molecules containing Li, Be, and B, the bond lengths for the eCEPPs are
distinctly underestimated, although by less than chemical accuracy.
Given the spatially large and highly polarizable [He] cores of these atoms,
it seems likely that the dominant source of this error is the incomplete
description of core relaxation provided by the CPP.

The negligible error for H$_2$ (of $1(3)\times10^{-4}$\AA) confirms
that most of the error for other molecules is due to the difficulty of
representing core-valence interaction with a pseudopotential rather
than limiting the reproduction of scattering properties to linear
order (all three pseudopotentials are equivalent for H).

The improvement in optimised geometries provided by the transition
metal eCEPPs is greater than that for the first row atoms.  This is
particularly apparent for TiO$_2$ for which the notable error in bond-angle
present for the other two pseudopotentials is completely removed.

Of all the transition metal molecules, the three containing Cu show
the largest underestimation of bond lengths, which is probably due to the
eventual emergence of an expected increase in pseudopotential error
with the increasing number of valence electrons.

The differences between molecular dissociation energies evaluated using CCSD(T)
for the eCEPP and AE Hamiltonians are shown in Fig.\ \ref{fig4}.
The figure also shows the dissociation energies previously reported
for the CEPP and TNDF pseudopotentials when available.

All eCEPP dissociation energies fall within chemical accuracy (of 43
meV) of the relativistic AE values, unlike the results for the TNDF
and CEPP pseudopotentials.  This improvement in accuracy is greater
than for optimised geometries, with the MAD for eCEPP
${\sim}1/10 \times$ that for TNDF pseudopotentials.
For all molecules considered, including H$_2$O$_2$ and H$_4$N$_2$,
the estimated basis set error is negligible.
Unlike the optimised geometries, errors in dissociation energies for molecules
containing Li, Be, or B are not unusually large and the CPP contribution to
energies is accurate.
The overall bias is small for the eCEPPs when compared with other
pseudopotentials, with an average deviation of 2 meV.

There is only a weak correlation between the errors for the TNDF and
CEPP potentials, and no correlation is apparent with the eCEPP errors.
No consistent trend is apparent for the eCEPPs except for a slow
increase in error with the number of valence electrons.
As for optimised geometries, the absence of core-valence
electron interaction for H$_2$ results in a negligible error (of
0.043(4) meV) for all three pseudopotentials.

Mean, absolute mean, and maximum deviations from AE results for the geometry
parameters and dissociation energies evaluated with the three pseudopotentials
types are shown in Table\ \ref{tab:1}.

\begin{table*}[t]
\begin{tabular}{lddd@{\hskip 24pt}ddd}                                        \\ \hline \hline
 & \multicolumn{3}{c}{ ($\times 10^{-3}$ \AA) } & \multicolumn{3}{c}{ (meV) } \\
Pseudopotential & 
\multicolumn{1}{c}{ $\overline{\Delta x_i}$ } &
\multicolumn{1}{c}{ $\overline{|\Delta x_i|}$ } &
\multicolumn{1}{c}{ Max$[\Delta x_i]$ } & 
\multicolumn{1}{c}{ $\overline{\Delta D_e}$ } &
\multicolumn{1}{c}{ $\overline{|\Delta D_e|}$ } &
\multicolumn{1}{c}{ Max$[\Delta D_e]$ }                                       \\ \hline
TNDF  &  1.2 & 4.3 &  31.1 &  13.65 &  117.05 &  431.70                       \\
CEPP  & -3.3 & 4.1 &  14.1 &  -8.89 &   30.72 &  99.11                        \\
eCEPP & -0.6 & 1.2 &  -7.5 &   1.67 &   11.04 &  42.02                        \\ \hline \hline
\end{tabular}
\caption{ \label{tab:1}
Summary of pseudopotential errors for TNDF pseudopotentials, CEPPs, and eCEPPs.
Values are for the mean, absolute mean, and maximum error taken over the
test set, denoted
$\overline{\Delta x_i}$, $\overline{|\Delta x_i|}$, and Max$[\Delta x_i]$
for geometry parameters, and
$\overline{\Delta D_e}$, $\overline{|\Delta D_e|}$, and Max$[\Delta D_e]$
for dissociation energies.
The test set of CEPPs does not include LiB, LiC, LiF, CuC, CuO, or CuF as
these molecules were not included in the previous work.
The test set for the TNDF pseudopotentials excludes the same molecules, 
as well as H$_2$O$_2$, CrF, and FeF due to lack of convergence.
}
\end{table*}

In order to assess the energetics absent from CCSD(T) the available monoxide
molecules are considered in more detail.
Table\ \ref{tab:2} show the equilibrium bond-lengths and dissociation energies
evaluated with all electrons present, and with core electrons represented by
the eCEPPs.
We compare this with the experimental data selected by Miliordos \emph{et al.}
for ScO--FeO\cite{Miliordos_SctoFe} and Harrison for CuO.\cite{Harrison_2000}
Results are also quoted for TiO\cite{Pan_2017} and CrO\cite{Chan_2012},
calculated using a composite method that corrects accurate CCSD(T)
energies to include spin-orbit coupling and higher-order correlation.

The disagreement between AE and eCEPP results is consistently negligible
compared with the significant disagreement with experimental results, for both
geometries and energies.

Bond lengths agree with experiment to within 0.01 \AA\ for all oxides except
CrO and MnO, with the largest underestimate of -0.016 \AA\ occurring for CrO.
Chan \emph{et al.} provide a CCSD(T) CrO bond length closer to experiment, but
this agreement is due to the small cc-pwCVTZ basis set and disappears with
a more complete basis.
This suggests that when an error in bond length occurs for these oxides it
is primarily due to missing energetics rather than pseudopotential error.

Only the ScO dissociation energy (for AE or eCEPP) agrees with experiment to
within 43 meV.
The maximum error, of -0.44 eV, occurs for FeO, with CCSD(T) overestimating
$D_e$ for ScO--VO and underestimating for CrO--CuO.
The composite dissociation energies for TiO and CrO are in much better
agreement with experiment, with most of the improvement provided by spin-orbit
corrections.
Higher-order correlation corrections are of secondary importance for these two
molecules.

So far we have not considered the accuracy with which the pseudopotentials
reproduce molecular potential energy surfaces beyond the minima, that is energy
differences considerably smaller than dissociation energies.
Hydrogen peroxide is of particular interest due to the relatively large error
in equilibrium spatial variables apparent in Fig.\ \ref{fig3}.
This is primarily due to the shallow variation of the total energy with
rotation of O-H$_2$ groups around the O-O axis, an internal rotation that plays
an important role in conformational analysis.\cite{Song_2005}

Geometries for H$_2$O$_2$ are optimised at the CCSD(T) level of theory using
the uncontracted aug-cc-pVTZ(-DK) basis sets (for AE calculations), with the
dihedral angle between O-H$_2$ groups, $\phi$, held constant.
Relaxed geometries are evaluated for several values of $\phi$, followed by an
extrapolation of the total energy to the CBS limit using $n=(TQ5)$.
Energy profiles evaluated with AEs present and using eCEPP pseudopotentials are
shown in Fig.\ \ref{fig5}.
Agreement is good, with a maximum pseudopotential and basis set error 
of 2.5 and 2.7 meV, respectively, and both occurring for $\phi=0$.

\begin{figure}[t]
\includegraphics[scale=1.00]{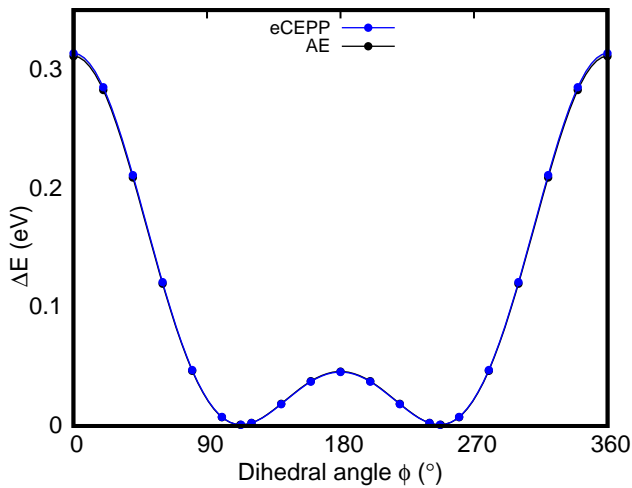}
\caption{ \label{fig5}
Energy profiles for the internal rotation of O-H$_2$ relative to each other
in H$_2$O$_2$. Results are shown from CCSD(T) calculations with all 
electrons present, and with core electrons represented by the eCEPPs.
The grey region indicates the error due to extrapolation to the CBS
limit for both AE and eCEPP calculations.
}
\end{figure}

Except for TiO and CrO, no attempt has been made to estimate the error due to
correlation neglected at the CCSD(T) level of theory.
While this error has been quantified as small when compared with basis
set errors for first-row molecules,\cite{Feller_2006} it is expected to
be significant for some molecules containing $3d$ transition metal
atoms.\cite{Jiang_2012}
However, this `missing correlation' is usually found to be between
valence electrons in molecules, rather than isolated atoms,
suggesting that the accuracy of the eCEPPs will be preserved 
for calculations including correlation beyond that present in CCSD(T).

\begin{table*}[t]
\begin{tabular}{lddddddd}									\\ \hline \hline
Method                       &
\multicolumn{1}{c}{\text{ScO}}  &
\multicolumn{1}{c}{\text{TiO}}  &
\multicolumn{1}{c}{\text{VO}}   &
\multicolumn{1}{c}{\text{CrO}}  &
\multicolumn{1}{c}{\text{MnO}}  &
\multicolumn{1}{c}{\text{FeO}}  &
\multicolumn{1}{c}{\text{CuO}}  								\\ \hline
          & \multicolumn{7}{c}{ $r_e$ (\AA) }							\\
eCEPP     & 1.669(6)  & 1.620(3)  & 1.584(5)  & 1.606(3) & 1.633(5)  & 1.616(2) & 1.730(2)	\\
AE        & 1.669(7)  & 1.619(3)  & 1.582(6)  & 1.605(5) & 1.633(5)  & 1.617(2) & 1.736(3)	\\
Composite & - & 1.619\footnotemark[1] & - & 1.6185\footnotemark[2] & - & - & -			\\
Expt.\footnotemark[3]
          & 1.6656    & 1.6203    & 1.589     & 1.6213   & 1.6477    & 1.6194   & 1.724		\\ \hline
          & \multicolumn{7}{c}{ $D_e$ (eV) }							\\
eCEPP     & 7.019(8)  & 7.097(10) & 6.765(10) & 4.540(5) & 3.712(9)  & 3.804(3) & 2.813(2)	\\
AE        & 7.024(11) & 7.080(11) & 6.749(10) & 4.521(4) & 3.696(12) & 3.791(5) & 2.776(3)	\\
Composite & - & 6.916\footnotemark[1] & - & 4.660\footnotemark[2] & - & - & - 			\\
Expt.\footnotemark[3]
          & 6.981(9)  & 6.93(7)   & 6.55(9)   & 4.63(7)  & 3.88(8)   & 4.235    & 2.928		\\
          &           &           &           & 4.80(9)  & 3.8(2)    &          &		\\ \hline \hline
\end{tabular}
\footnotetext[1]{ From [\onlinecite{Pan_2017}]. Bond lengths are optimised using CCSD(T),
                  and dissociation energies are CCSD(T) corrected to include
                  spin-orbit coupling and higher-order correlation.}
\footnotetext[2]{ From [\onlinecite{Chan_2012}]. As [\onlinecite{Pan_2017}].}
\footnotetext[3]{ Sum of experimental atomisation energies and zero-point vibrational energies.
                  Data as cited by [\onlinecite{Miliordos_SctoFe}]
                  for Sc--Fe, and [\onlinecite{Harrison_2000}] for Cu. }
\caption{ \label{tab:2}
 Bond lengths ($r_e$) and dissociation energies ($D_e$) for monoxides of Sc--Fe
 and Cu.
 Results are shown evaluated using CCSD(T) with all electrons present, and with
 eCEPPs.
 Experimental values from the literature are shown for all oxides,
 and previously published composite results for TiO and CrO are evaluated by
 correcting CCSD(T) energies by including spin-orbit coupling and higher-order
 correlation effects (with geometries optimised using CCSD(T) only).
}
\end{table*}

\subsection{Comparison of errors for a theory hierarchy\label{sec:errors}}

An important application of the eCEPPs is expected to be to
QMC methods for bulk systems, particularly Diffusion Monte Carlo (DMC).
DMC requires a fermionic nodal surface to be supplied, and for bulk systems
this is normally approximated as the nodal surface of a determinant constructed
from orbitals resulting from a KS-DFT calculation with a plane-wave basis.

In light of this we begin with an AE CCSD(T), then quantify and compare the
error introduced by each approximation involved in an eCEPP KS-DFT
calculation with a plane-wave basis set.
The purpose of this analysis is to compare the relative size of the various
components of the error in the final calculation.

If the error due to the eCEPP alone is small when compared with 
other errors then it may be expected that the contribution of the
eCEPP to the error in the approximate nodal surface is also small.
While this measure of the accuracy of the nodal surface is not rigorous
(an exact exchange-correlation functional would not provide an exact nodal
surface, and in principle, an inaccurate pseudopotential or
exchange-correlation functional could provide a more accurate nodal surface),
it is physically reasonable that less accurate KS-DFT molecular properties 
will correspond to a less accurate nodal surface.

A separation of errors naturally arises by considering the application of 
several levels of approximation.
The most accurate description considered is AE CCSD(T) with a Gaussian basis.
The next level of approximation is to replace atomic cores with the full eCEPP.
We then increase the level of approximation progressively by removing the CPP
component of the eCEPP, followed by replacing CCSD(T) with KS-DFT, and finally
by replacing the full eCEPP with a Kleinman-Bylander\cite{Kleinman_1982}
representation (of the eCEPP) together with a plane-wave basis.
Both KS-DFT calculations are performed using the PBE exchange-correlation
functional.\cite{Perdew_1996}

The KB pseudopotential representation of the eCEPP is given by
\begin{equation}
\label{kb}
\hat{V} = V^{pp}_{l_{max}} + \sum_l \sum_{ij}
\left| \delta V_l \phi_{li} \right> B_{l,ij} \left< \phi_{lj} \delta V_l \right|,
\end{equation}
with $\delta V_l = V^{pp}_l - V^{pp}_{l_{max}}$.
The $\phi_{li}$ functions used for each projector are supplied
for each channel of each pseudopotential, and the matrices for each
channel are given by
$B_{l,ij}= \left< \phi_{li} | \delta V_l | \phi_{lj} \right>^{-1}$.
This separable form is less accurate than the semi-local spherical
harmonic projector representation used to define the eCEPPs.
However, the KB representation provides a considerable reduction
in computational cost, and the additional error introduced is often small.

Three projection orbitals are used for first row atoms, and six projection
orbitals for the $3d$ transition metal atoms.
For all atoms the local channel is taken to be that with the largest $l$
value, $l_{max}$.
In each case orbitals are taken from the neutral atom, and supplemented
by orbitals from the lowest-energy excited states that provide
projectors for all pseudopotential channels.\cite{Supplemental}
For example, for Ti the ground state provides projectors from the $3s$, $4s$,
$3p$ and $3d$ orbitals, to which we add $4p$ and $4f$ projectors from the
[Ar]$3d^24s4p$ and [Ar]$3d^24f$ states, respectively.

Note that projectors for $l=l_{max}$ are evaluated but unused.
These are provided to allow flexibility in the choice of the local channel
of the KB representation, should it be required.
All projectors are generated using the PBE functional and the atomic KS-DFT
code contained in the \textsc{Quantum Espresso} 
package.\cite{Giannozzi_espresso_2009}

Errors are assessed for the subset of diatomic molecules present in the full
test set, excluding the excited TiN states.
Bond lengths are relaxed and dissociation energies evaluated for each of these
molecules, and at each level of theory.

All electron and eCEPP results with the CPP are identical
to those included in Sec.\ \ref{sec:molecules}.
Dissociation energies with the CPP removed are evaluated with the same data but
excluding CPP corrections.

Results for KS-DFT without the KB representation are generated
using the uncontracted aug-pp-pV$5$Z Gaussian basis sets.
Extrapolation is not required as exponential convergence results
from correlation being contained in the exchange-correlation functional,
rather than the wave function itself.
Estimated basis set error is $<$8 meV, from differences between
dissociation energies evaluated using the $n=(Q5)$ basis sets.

For KS-DFT with a plane-wave basis set the isolated molecule is modelled by
a single molecule in a $(15$\AA$)^3$ periodic unit cell, and
using the \textsc{CASTEP}\cite{castep} plane-wave pseudopotential code (with a
small modification to allow for more than one projector per pseudopotential
channel).
Calculations for all atoms and molecules are performed using the
plane-wave basis set defined by an energy cutoff of 150 Ha, 
resulting in a basis set error of $<$9 meV.
The maximum error occurs for CuF due to the number of valence electrons present and
the depth of the Cu and F potentials.
The same accuracy could be achieved with considerably smaller basis sets for
most of the molecules in our test set.

The total error arising from all approximations taken together is
\begin{eqnarray}
\label{err_components1}
\Delta                 & = & D_e^{\text{pp-cpp,DFT+KB}} - D_e^{\text{AE,CCSD(T)}}      \nonumber \\
                       & = & \Delta_{\text{pp}} + \Delta_{\text{cpp}} + \Delta_{\text{DFT}} + \Delta_{\text{KB}}.
\end{eqnarray}
Components of the total error due to pseudopotentials, the absent CPP, 
density functional theory (with PBE), and the KB representation are given by
\begin{eqnarray}
\label{err_components2}
\Delta_{\text{pp}}     & = & D_e^{\text{pp,CCSD(T)}}     - D_e^{\text{AE,CCSD(T)}}     \nonumber \\
\Delta_{\text{cpp}}    & = & D_e^{\text{pp-cpp,CCSD(T)}} - D_e^{\text{pp,CCSD(T)}}     \nonumber \\
\Delta_{\text{DFT}}    & = & D_e^{\text{pp-cpp,DFT}}     - D_e^{\text{pp-cpp,CCSD(T)}} \nonumber \\
\Delta_{\text{KB}}     & = & D_e^{\text{pp-cpp,DFT+KB}}  - D_e^{\text{pp-cpp,DFT}}     ,
\end{eqnarray}
respectively.
The superscripts for each dissociation energy in Eq.\ (\ref{err_components2})
describe the level of approximation.
Superscripts `AE', `pp', and `pp-cpp' denote calculations including all
electrons, describing core electrons using an eCEPP, and describing core electrons
using an eCEPP with core polarisation removed.
Superscripts `CCSD(T)', `DFT', and `DFT+KB' denote the use of CCSD(T),
KS-DFT, and KS-DFT with a KB representation of the eCEPPs.

\begin{figure*}[t]
\includegraphics[scale=1.00]{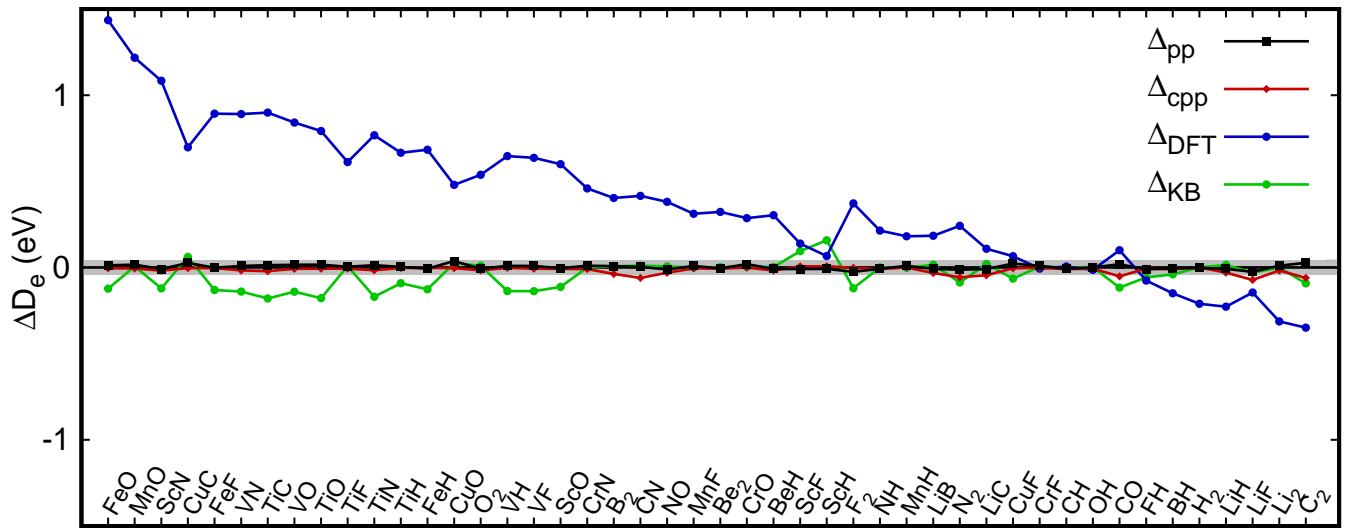}
\caption{ \label{fig6}
A breakdown of the errors in dissociation energies for diatomic molecules.
Errors are shown arising from the eCEPPs themselves ($\Delta_{pp}$),
from excluding core polarisation potentials ($\Delta_{cpp}$),
from employing KS-DFT (with PBE exchange-correlation) ($\Delta_{DFT}$), and
from using a Kleinman-Bylander representation for the eCEPPs ($\Delta_{KB}$).
Differences within the grey region are less than 43 meV.
}
\end{figure*}

The error breakdown is shown in Fig.\ \ref{fig6}.
The overall trend is for KS-DFT errors to be dominant when compared
with the three errors associated with the pseudopotential.
The negligible error that arises for the CrF, CH, and OH molecules 
is unlikely to be more than fortuitous.
The largest KS-DFT errors arise for molecules containing transition
metal atoms, which is unsurprising given the strong correlation and highly
inhomogeneous charge densities for these molecules.

Errors due to the absence of a CPP are almost negligible compared to the KS-DFT
error.
Molecules containing Li, with the most polarizable core, show the largest
contribution from the CPP, although the net polarizability stands out as
particularly large for CN, N$_2$, CO and C$_2$.

The KB representation results in a relatively small error.
No overall pattern is apparent except that the KB representation is
significantly less accurate for molecules containing transition metal
atoms, and the error is greatest in TiC. 
This is as expected, since the non-local and local channels differ the most
for these atoms, with the eCEPPs composed a deep $d$ channels and relatively
shallow $s$, $p$, and $f$ channels.

Physically unrealistic `ghost' eigenstates can arise when the KB representation
is used.\cite{Gonze_1991}
Agreement between KS-DFT energy differences evaluated with and without the KB
representation (see Fig.\ \ref{fig6}) provide a useful check that such ghost
states do not occur for eCEPPs with the local channel chosen as $l=l_{max}$,
and this is further confirmed by good agreement between total energies (not
shown).
This absence of ghost states for the $3d$ transition metals is very unlike the
TNDF pseudopotentials, \cite{Drummond_2016} and can be ascribed to the presence
of a physically realistic eCEPP channel for $l\ge3$.

The breakdown of errors in bond lengths (not shown) shows similar behaviour,
except that the error due to the absence of a CPP is comparable to the KS-DFT error
for molecules containing Li, Be, and B, and KB error is dominant for ScH and ScF.
Unlike the dissociation energies, bond lengths for CrF, CH, and OH are not
notably accurate.

\subsection{Optimised basis sets \label{sec:eCEPP_basis}}

\begin{figure*}[t]
\includegraphics[scale=1.00]{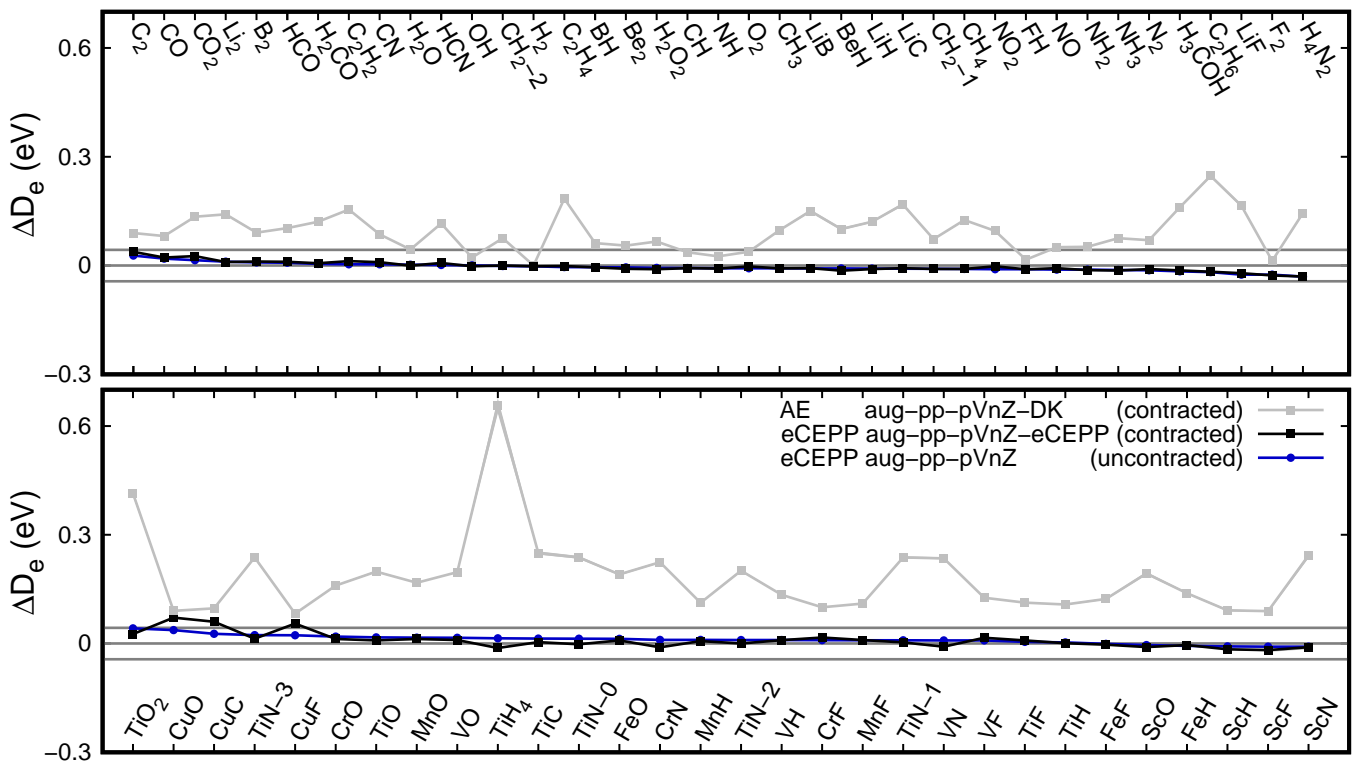}
\caption{ \label{fig7} 
Comparison of errors in CCSD(T) dissociation energies for different basis sets,
evaluated using both AE and eCEPPs.
Errors are estimated as the deviation from the AE $D_e$ provided by CCSD(T) with
uncontracted aug-pp-pV$n$Z-DK basis sets.
Blue shows the errors in eCEPP energies evaluated using the uncontracted
aug-pp-pV$n$Z basis sets.
Black shows the errors in eCEPP energies evaluated using the contracted basis
set optimised for eCEPPs, aug-pp-pV$n$Z-eCEPP.
Grey shows the error for AE dissociation energies evaluated using the contracted
aug-pp-pV$n$Z-DK basis sets.
Grey lines at $\pm$43 meV contain all data points that
reproduce AE energies to within chemical accuracy.
}
\end{figure*}

Contracted correlation consistent Gaussian basis sets are well
established to be computationally efficient for correlated electron
calculations, providing controlled convergence behaviour that allows 
accurate extrapolation to the CBS limit.

The eCEPP pseudo atom wave-functions are very different from AE wave-functions
in the core region, hence such a basis set must be reconstructed for each
eCEPP.
We name basis sets optimised for the eCEPPs `aug-cc-pV$n$Z-eCEPP'.
As for the AE case, these are constructed by optimising a Gaussian basis within
HF, systematically introducing correlation consistent components of the basis
at higher angular momentum quantum numbers, and augmenting
the basis with further Gaussians optimised to accurately represent
the more diffuse anionic state.

The basis sets are optimised using the procedure developed for 
first row atoms by Xu \emph{et al.},\cite{Xu_2013}
and the $3d$ transition metal approach of Balabanov
\emph{et al.},\cite{basisDK} adapted to cores described by
pseudopotentials.\cite{Supplemental}

In the presence of pseudopotentials, valence states become smoother and core
states are absent, hence fewer Gaussians and contractions are required than
for the AE case.
This results in a significant improvement in computational efficiency,
and the average CCSD(T) run-time is $\sim1/5\times$ that for
uncontracted standard basis sets.
This speedup should increase with molecular size.

Figure\ \ref{fig7} shows the errors in eCEPP CCSD(T) dissociation energies
evaluated using both the \emph{uncontracted} standard aug-cc-pV$n$Z basis set,
and the new \emph{contracted} aug-cc-pV$n$Z-eCEPP basis set.
The error is quantified as the deviation from the baseline of AE CCSD(T)
energies evaluated using the \emph{uncontracted} standard aug-cc-pV$n$Z-DK basis
set.

For molecules containing first row atoms the new contracted basis
sets reproduce the uncontracted aug-cc-pV$n$Z basis set errors accurately.
Molecules containing $3d$ transition metal atoms show a less consistent
agreement.
However, the additional error is small and marginally greater than
chemical accuracy only for the three molecules containing copper.
For most cases the error due to the contracted aug-cc-pV$n$Z-eCEPP basis
is similar or less than that due to the pseudopotential itself.

In Fig.\ \ref{fig7} we also show the error introduced to AE CCSD(T) by using
\emph{contracted} standard aug-cc-pV$n$Z-DK basis sets.
This is quantified as the difference between dissociation
energies evaluated using the \emph{contracted} and \emph{uncontracted} 
aug-cc-pV$n$Z-DK basis sets, with the latter being the more complete basis,
which provides a useful scale for assessing the accuracy of the
aug-cc-pV$n$Z-eCEPP basis sets.

The basis set error introduced by contraction is significantly greater for
the AE calculations than for the eCEPP calculations (except for H$_2$ for which
both are negligible), especially for the molecules containing $3d$ transition
metal atoms.
We conclude that the construction of aug-cc-pV$n$Z-eCEPP basis sets and
application with the eCEPPs is successful, with an impressively low error for
the test set considered.

Finally, we note that the relatively large error in the relativistic AE
dissociation energy calculated with the contracted aug-cc-pV$n$Z-DK basis
sets is not unexpected, and is not a deficiency of these basis sets
\emph{per se}.
These standard contractions are constructed to represent correlations
between electrons outside of a [He] core for Li--F, and outside of an
[Ar] core for Sc--Cu, so do not describe correlation between these cores and
valence electrons well, despite the CCSD(T) active space including excitation
of all electrons.

This explanation is consistent with the very small error achieved for H$_2$,
with the larger errors for the $3d$ transition metal molecules, and the
accuracy of the aug-cc-pwCV$n$Z basis sets\cite{basisDK} for which contractions
include core-valence correlation effects.

\section{Conclusions \label{sec:conclusions}}
The paradigms of shape and energy consistency in the definition of electronic
pseudopotentials are successfully and consistently combined to generate
accurate effective potentials.
The resulting eCEPPs reproduce core electron scattering properties to first order,
reproduce atomic excitation and ionisation energies,
and partially reproduce core relaxation effects via a dipole polarisation potential.
This approach is an extension of the CEPPs (that reproduce only
core electron scattering and core polarisation).
Electron correlation is included throughout, no recourse is made to independent
electron orbitals, and the generation process is \emph{ab initio}.

Comparing AE and eCEPPs results for relaxed geometries and dissociation energies,
and for a test set of 65 molecules (in 69 states) at the CCSD(T) level of theory,
demonstrates a reduction in error by ${\sim}1/4\times$ for geometry parameters,
and by ${\sim}1/10\times$ in dissociation energies when compared with
Hartree-Fock pseudopotentials.

When compared with the previously published CEPPs, the eCEPPs 
introduce an interpolation of the effect of the core on several
atomic states in addition to the extrapolation of core scattering.
This extension is significant, reducing errors in geometry parameters and dissociation 
energies by a factor of ${\sim}1/4\times$ and ${\sim}2/5\times$.

The eCEPPs will primarily be useful for correlated wave-function methods,
particularly for a CCSD(T) description of molecular systems,
and for the application of QMC methods to extended systems.
For both cases pseudopotentials are often unavoidable and
the eCEPPs provided here are validated by small errors for a reasonably large
test set.

The improved accuracy is particularly important when $3d$ transition metal
atoms are present as the eCEPP error is sufficiently small to
resolve the effect of spin-orbit coupling and aspects of electron-electron
correlation often absent in \emph{ab initio} many-body methods.

Further work naturally suggests itself.
It would be useful and desirable to extend the eCEPP generation process to more
atomic types.
Such an extension is however limited by the size and stability 
of MCHF and CCSD(T) calculations, and the absence of 
correlation consistent basis sets for heavier atoms.
However, completing the H--Kr set of atoms can be expected
to be straightforward.

It is perhaps more interesting to consider the accuracy of the eCEPPs,
and how much further improvement is possible.
In principle, the accuracy could be improved by extending the reproduction of
core scattering properties to higher order and many valence electrons.
As discussed in the introduction, implementing this is non-trivial
as it will result in a many-body effective potential.

Another possible strategy would be to extend the reproduction of 
atomic ionisation and excitation energies to valence charge densities outside
of the core region.
Given the success of the eCEPP in reproducing these densities for TiO,
such an extension appears to be unnecessary.

The most direct approach to improving the eCEPP accuracy lies
in the choice of excited/ionic states for which energy consistency is
enforced.
Here we have considered the atomic states with the lowest energy.
A more detailed consideration of the molecular bonds each atom is
prone to form should, in principle, allow a selection of atomic states which
are more appropriate for each atom.
This could also be extended to molecular geometries and dissociation energies.
The primary drawback of this approach (and why it was not used here) is
that the pseudopotential generation process would become more biased.

A simple, and possibly effective, approach to reducing pseudopotential error
would be a more careful choice of core radii.
These radii parameterise a balance between errors due to defining the eCEPP
with a finite core charge density outside of the core region, and errors due
to replacing the valence charge density with a pseudo density within the core
region.
Optimised core radii, different from those used in this paper, may well 
decrease errors significantly.
This is likely to be most influential for the lightest atoms with the
most diffuse cores.

The most important issue that has not been addressed is the accuracy of the
eCEPPs when used in QMC calculations.
CCSD(T) results are accurate, and DMC is known to provide a similar
or better accuracy than CCSD(T).
However, DMC involves a further `pseudopotential localisation'
approximation that expresses the semi-local one-body pseudopotential(s)
as a local many-body potential via a trial wave-function.
The resulting error is second order in the accuracy of the trial 
wave-function.\cite{Foulkes_QMC_review,Casino_reference}
A measure of this error would be desirable, and the molecules considered here
are useful candidates for such tests.

\section*{Supplementary Material}
See supplementary material for details of CCSD(T) calculations, atomic and
molecular states, KB projectors, eCEPP basis set optimisation, and the eCEPPs
themselves.

\begin{acknowledgments}
  R.J.N.\ and J.R.T.\ acknowledge financial support from the
  Engineering and Physical Sciences Research Council (EPSRC) of the
  U.K.\ [EP/J017639/1].
\end{acknowledgments}


\end{document}


\title{Shape and Energy Consistent Pseudopotential for Correlated Electrons:
       Supplemental Material}

\author{J. R. Trail}
\affiliation{Theory of Condensed Matter Group, Cavendish Laboratory, J
  J Thomson Avenue, Cambridge CB3 0HE, United Kingdom}

\author{R. J. Needs}
\affiliation{Theory of Condensed Matter Group, Cavendish Laboratory, J
  J Thomson Avenue, Cambridge CB3 0HE, United Kingdom}

\date{\today}

\maketitle

\subsection{Basis sets and complete basis set limit for CCSD(T) \label{sec:ccsd}}

Coupled-cluster calculations with single, double, and perturbative
triple excitations are used to generate the eCEPPs, and to test their accuracy
by comparison of pseudopotential and AE results.
The standard all-electron augmented correlation-consistent
relativistic basis sets aug-cc-pV$n$Z-DK\cite{basis,basisDK} are used for
AE calculations, with relativistic effects included by employing the
second order Douglas-Kroll-Hess Hamiltonian.
For the eCEPPs relativistic effects are implicit in
the pseudopotential, hence a non-relativistic Hamiltonian is used with the
aug-cc-pV$n$Z\cite{basis} basis.
(The parameter $n$ is the number of correlating functions included in the
basis.)

Both of these basis sets are available with contractions.
The contractions are generated for neutral AE atoms, so are inaccurate for
very ionic AE states and for all pseudo-atom states.
Consequently we use uncontracted basis sets for both AE and
pseudopotential calculations.

The CCSD(T) active space is defined to include core-valence correlation by
allowing excitations of all electrons.

Differences between total energies, such as dissociation or ionisation
energies, are provided in the CBS limit using
extrapolation, with an accompanying estimate of the extrapolation error.
Two-point extrapolation, together with an estimated error, is performed
using \cite{Trail_2013_pseudopotentials,Trail_2015_pseudopotentials}
\begin{eqnarray}
\label{extrap}
\Delta E[n] &=& a_1 + b_1 e^{-n}   \nonumber \\
\Delta E[n] &=& a_2 + b_2 n^{-3}   \nonumber \\
\Delta E^{est} &=&
\frac{a_1+a_2}{2}
 + \Delta E^{cpp}[T] - \Delta E[T] \pm \frac{|a_1-a_2|}{2}
\end{eqnarray}
where $\Delta E[n]$ is the difference between CCSD(T) total energy provided by
the aug-cc-pV$n$Z(-DK) basis set.
Estimates of the CBS energy and error are taken as half the sum and half the
absolute difference of parameters $a_1$ and $a_2$.
A correction is included that introduces CPP effects using total energy
differences with ($\Delta E^{cpp}[T]$) and without ($\Delta E[T]$) the CPP part
of the Hamiltonian, and evaluated with the $n=T$ basis set.
This CPP correction is zero for the AE atoms.

Total CBS energies are generated from calculations using the $n=(TQ5)$
basis sets (two-point extrapolation using $n=(Q5)$
together with the CPP corrections obtained using $n=(T)$ when appropriate).
This extrapolation is a simplified version of that used by
Feller \emph{et al.},\cite{Feller_2010} modified to include CPP effects
when required.

State-averaged Hartree-Fock is used to enforce symmetries for transition metal
atoms and diatomic molecules, which is required due to the \textsc{MOLPRO}
implementation not including symmetry involving continuous rotations.
All CCSD(T) calculations are performed with orbitals provided by restricted
open shell Hartree-Fock (ROHF) calculations.

Geometry optimisation is performed for each molecule by minimisation of
the AE and pseudopotential CCSD(T) energy with no extrapolation to the CBS
limit, and uncontracted basis sets.

Bond lengths, bond--angles and dihedral angles that completely characterise the
geometry of a molecules of known symmetry provide the geometry parameters used.
Each estimated parameter value, $x_i^{est}$ is given by
\begin{equation}
\label{geom}
x_i^{est} = x_i[n] + x^{cpp}_i[T] - x_i[T] \pm |x_i[n] - x_i[n-1]|/2,
\end{equation}
where $x_i[n]$ are the geometry parameters arising from optimisation
with an aug-cc-pV$n$Z(-DK) basis set.
The CPP correction, $x^{cpp}_i[T] - x_i[T]$, is zero for AE atoms,
whereas for pseudo-atoms the $x^{cpp}_i[T]$ are the optimum
parameters that occur for the CPP included in the Hamiltonian.
The estimated error is half the absolute difference between optimum
parameters for the aug-cc-pV$n$Z(-DK) and aug-cc-pV($n-1$)Z(-DK) basis set.

For most of the molecules the geometry estimate is generated using the
$n=(TQ5)$ basis sets (optimum geometries from the $n=5$ basis, the CPP
correction from the $n=T$ basis, and basis set error from the $n=(Q5)$
basis sets).
Geometries are estimated using $n=(TTQ)$ for the molecules containing $3d$
transition metal atoms and the larger first-row molecules (C$_2$H$_2$, H$_2$CO,
H$_2$O$_2$, C$_2$H$_4$, H$_3$COH, H$_4$N$_2$, and C$_2$H$_6$).

\subsection{Ion and neutral atomic states used to define energy consistency \label{sec:sec:energy_consistency}}

\begin{table*}[t]
\begin{tabular}{lll@{\hskip 18pt}ll@{\hskip 18pt}ll@{\hskip 18pt}ll@{\hskip 18pt}ll@{\hskip 18pt}ll}
    \hline \hline
  & \multicolumn{12}{c}{Ionisation} \\
    \multicolumn{1}{l}{Atom} & & $+1$ & & +0 & & +1 & & +2 & & +3 & & +4 \\ \hline
H   & $^1S$ & $1s^2$	     & $^2S$ & $1s$   	        & -     & -		    &       &                &       &            & & \\
    &       &		     & $^2P$ & $1p$             & 	&	            &       &                &       &            & & \\
    &       &		     & $^2D$ & $1d$             & 	&	            &       &                &       &            & & \\ \hline
Li  & $^1S$ & [He]$2s^2$     & $^2S$ & [He]$2s$         & $^1S$ & -	            &       &                &       &            & & \\
    &       &		     & $^2P$ & [He]$2p$         &	&	            &       &                &       &            & & \\
    &       &		     & $^2D$ & [He]$3d$         &	&	            &       &                &       &            & & \\
Be  & $^2S$ & [He]$2s^22p$   & $^1S$ & [He]$2s^2$       & $^2P$ & [He]$2p$          & $^1S$ & -		     &       &            & & \\
    &       &		     & $^3P$ & [He]$2s2p$       &	&	            &       &                &       &            & & \\
B   & $^3P$ & [He]$2s^22p^2$ & $^2P$ & [He]$2s^22p$     & $^1S$ & [He]$2s^2$	    &       &  	 	     & $^1S$ & -          & & \\
    &       &		     & $^4P$ & [He]$2s2p^2$     &	&	            &       &                &       &            & & \\
C   & $^4S$ & [He]$2s^22p^3$ & $^3P$ & [He]$2s^22p^2$   & $^2P$ & [He]$2s^22p$      & $^1S$ & [He]$2s^2$     &       &            & & \\
    &       &		     & $^5S$ & [He]$2s2p^3$     & $^2D$ & [He]$2s^23d$      &       &                &       &            & & \\
    &       &		     & $^5P$ & [He]$2s2p^23s$   &	&	            &       &                &       &            & & \\
N   & $^3P$ & [He]$2s^22p^4$ & $^4S$ & [He]$2s^22p^3$   & $^3P$ & [He]$2s^22p^2$    & $^2P$ & [He]$2s^22p$   & $^1S$ & [He]$2s^2$ & & \\
    &       &		     & $^4P$ & [He]$2s2p^4$     & $^3F$ & [He]$2s^22p   3d$ &       &                &       &            & & \\
O   & $^2P$ & [He]$2s^22p^5$ & $^3P$ & [He]$2s^22p^4$   & $^4S$ & [He]$2s^22p^3$    & $^3P$ & [He]$2s^22p^2$ &       &            & & \\
    &       &		     & $^5S$ & [He]$2s^22p^33s$ & $^4P$ & [He]$2s  2p^4$    &       &                &       &            & & \\
    &       &		     & $^5P$ & [He]$2s^22p^33p$ &	&	            &       &                &       &            & & \\
    &       &		     & $^5D$ & [He]$2s^22p^33d$ &	&	            &       &                &       &            & & \\
F   & $^1S$ & [He]$2s^22p^6$ & $^2P$ & [He]$2s^22p^5$   & $^3P$ & [He]$2s^22p^4$    &       &                &       &            & & \\
    &       &		     & $^4P$ & [He]$2s^22p^43s$ & $^3P$ & [He]$2s  2p^5$    &       &                &       &            & & \\
    &       &		     & $^4P$ & [He]$2s^22p^43p$ &	&	            &       &                &       &            & & \\ \hline
Sc  & $^3P$ & [Ar]$4s^24p3d$    & $^2D$ & [Ar]$4s^23d$    & $^3D$ & [Ar]$4s3d$    & $^2D$ & [Ar]$3d$     & $^1S$ & [Ar]	        &       &      \\
    & $^4F$ & [Ar]$4s3d^2$      & $^3F$ & [Ar]$3d^2$      & $^2S$ & [Ar]$4s$      &       &	         &       &              &       &      \\
Ti  & $^4F$ & [Ar]$4s^23d^3$    & $^3F$ & [Ar]$4s^23d^2$  & $^4F$ & [Ar]$4s3d^2$  & $^3F$ & [Ar]$3d^2$   & $^2D$ & [Ar]$3d$     & $^1S$ & [Ar] \\
    &       &		        & $^5F$ & [Ar]$4s3d^3$    & $^4F$ & [Ar]$3d^3$    & $^3D$ & [Ar]$4p3d$   &       &              &       &      \\
V   & $^5D$ & [Ar]$4s^23d^4$    & $^4F$ & [Ar]$4s^23d^3$  & $^5D$ & [Ar]$3d^4$    & $^4F$ & [Ar]$3d^3$   & $^3F$ & [Ar]$3d^2$   & $^2D$ & [Ar]$3d$   \\
    &       &		        & $^6D$ & [Ar]$4s3d^4$    &       &               & $^4G$ & [Ar]$4p3d^2$ &       &              &       &      \\
Cr  & $^6S$ & [Ar]$4s^23d^5$    & $^7S$ & [Ar]$4s3d^5$    & $^6S$ & [Ar]$3d^5$    & $^5D$ & [Ar]$3d^4$   & $^4F$ & [Ar]$3d^3$   &       &      \\
    &       &		        & $^5D$ & [Ar]$4s^23d^4$  & $^6F$ & [Ar]$4p3d^4$  & $^5F$ & [Ar]$4s3d^3$ &       &              &       &      \\
Mn  &       &   	        & $^6S$ & [Ar]$4s^23d^5$  & $^7S$ & [Ar]$4s3d^5$  & $^6S$ & [Ar]$3d^5$   & $^5D$ & [Ar]$3d^4$   & $^4F$ & [Ar]$3d^3$ \\
    &       &		        & $^6D$ & [Ar]$4s3d^6$    &       &               & $^7P$ & [Ar]$4p3d^5$ & $^6D$ & [Ar]$4s3d^4$ &       &      \\
Fe  & $^4F$ & [Ar]$4s^23d^7$    & $^5D$ & [Ar]$4s^23d^6$  & $^6D$ & [Ar]$4s3d^6$  & $^5D$ & [Ar]$3d^6$   & $^6S$ & [Ar]$3d^5$   &       &      \\
    &       &		        & $^5F$ & [Ar]$4s3d^7$    & $^6D$ & [Ar]$4p3d^6$  & $^7S$ & [Ar]$4s3d^5$ &       &              &       &      \\
Cu  & $^1S$ & [Ar]$4s^23d^{10}$ & $^2S$ & [Ar]$4s3d^{10}$ & $^1S$ & [Ar]$3d^{10}$ & $^2D$ & [Ar]$3d^9$   & $^3F$ & [Ar]$3d^8$   &       &      \\
    &       &		        & $^2D$ & [Ar]$4s^23d^9$  & $^3D$ & [Ar]$4s3d^9$  & $^4F$ & [Ar]$4s3d^8$ &       &              &       &      \\
\hline \hline
\end{tabular}
\caption{ \label{tab:1}
 Atomic states used for energy-consistency condition in eCEPP generation.
 Hydrogen has no core, Li--F have a [He] core, and for Sc--Fe and Cu the core is [Ne].
 For Sc--Fe and Cu the $3s^23p^6$ is left implicit and must be included in each state.
 Pseudo-atoms with no valence electrons and an energy of zero are denoted `-'.
}
\end{table*}

For H--B we include the neutral atomic ground states, one anionic
ground state, and all ionic ground states allowed for the given
core.
The lowest energy ionic state corresponds to a pseudo-atom with no bound
electrons and a total energy of zero.

For C--F we include neutral atomic ground states, one anionic state,
and states ionised to $+2$, $+3$, $+2$, $+1$ respectively.
While we do not include all possible states for these atoms, we
consistently include the lowest energy excited states for the neutral atoms and
$+1$ ion.

For Sc--Fe and Cu we select states similarly, including anion, neutral,
and ionic ground states, together with 3 excited states for each atom.
States ionised to $+3$ are used for Sc, Cr, Fe, and Cu and to $+4$ for
Ti, V, Mn, with an excited anionic state included for Sc and the
anionic ground state excluded for Mn (it is not stable).

Selected states for each atom are shown in Table\ \ref{tab:1}.

\subsection{Molecular test set \label{sec:test_set}}

\begin{table*}[t]
\begin{tabular}{ll@{\hskip 24pt}ll@{\hskip 24pt}ll@{\hskip 24pt}ll} \hline \hline
\multicolumn{8}{c}{Molecules and terms} \\ \hline
H$_2$      &  $^1 \Sigma_g$ & CH$_2$-1   &  $^1 A_1     $ & ScH        &  $^1 \Sigma  $ & VH         &  $^5 \Delta  $ \\
LiH        &  $^1 \Sigma  $ & CH$_2$-2   &  $^3 B_1     $ & ScN        &  $^1 \Sigma  $ & VN         &  $^3 \Delta  $ \\
BeH        &  $^2 \Sigma  $ & NH$_2$     &  $^2 B_1     $ & ScO        &  $^2 \Sigma  $ & VO         &  $^4 \Sigma  $ \\
BH         &  $^1 \Sigma  $ & H$_2$O     &  $^1 A_1     $ & ScF        &  $^1 \Sigma  $ & VF         &  $^5 \Pi     $ \\
CH         &  $^2 \Sigma  $ & HCN        &  $^1 \Sigma  $ & TiH        &  $^4 \Phi    $ & CrN        &  $^4 \Sigma  $ \\
NH         &  $^3 \Sigma  $ & HCO        &  $^2 A'      $ & TiC        &  $^3 \Sigma  $ & CrO        &  $^5 \Phi    $ \\
OH         &  $^2 \Sigma  $ & CO$_2$     &  $^1 \Sigma_g$ & TiN-0      &  $^2 \Sigma  $ & CrF        &  $^6 \Sigma  $ \\
FH         &  $^1 \Sigma  $ & NO$_2$     &  $^2 A_1     $ & TiN-1      &  $^2 \Delta  $ & MnH        &  $^7 \Sigma  $ \\
Li$_2$     &  $^1 \Sigma_g$ & CH$_3$     &  $^2 A_2''   $ & TiN-2      &  $^4 \Delta  $ & MnO        &  $^6 \Sigma  $ \\
Be$_2$     &  $^1 \Sigma_g$ & NH$_3$     &  $^1 A_1     $ & TiN-3      &  $^2 \Pi     $ & MnF        &  $^7 \Sigma  $ \\
B$_2$      &  $^3 \Sigma_g$ & C$_2$H$_2$ &  $^1 \Sigma_g$ & TiO        &  $^3 \Delta  $ & FeH        &  $^4 \Delta  $ \\
C$_2$      &  $^1 \Sigma_g$ & H$_2$CO    &  $^1 A_1     $ & TiF        &  $^4 \Phi    $ & FeO        &  $^5 \Delta  $ \\
N$_2$      &  $^1 \Sigma_g$ & H$_2$O$_2$ &  $^1 A       $ & TiH$_4$    &  $^1 A_1     $ & FeF        &  $^6 \Delta  $ \\
O$_2$      &  $^3 \Sigma_g$ & CH$_4$     &  $^1 A_1     $ & TiO$_2$    &  $^1 A_1     $ & CuC        &  $^2 \Pi     $ \\
F$_2$      &  $^1 \Sigma_g$ & C$_2$H$_4$ &  $^1 A_g     $ &            &                & CuO        &  $^2 \Pi     $ \\
CN         &  $^2 \Sigma  $ & H$_3$COH   &  $^1 A'      $ &            &                & CuF        &  $^1 \Sigma  $ \\
CO         &  $^1 \Sigma  $ & H$_4$N$_2$ &  $^1 A       $ &            &                &            &                \\
NO         &  $^2 \Sigma  $ & C$_2$H$_6$ &  $^1 A_{1g}  $ &            &                &            &                \\
LiB        &  $^3 \Sigma  $ &                             &            &                &            &                \\
LiC        &  $^4 \Sigma  $ &                             &            &                &            &                \\
LiF        &  $^1 \Sigma  $ &                             &            &                &            &                \\ \hline \hline
\end{tabular}
\caption{ \label{tab:2}
 Symmetries of the test set of molecules and states used to compare optimum geometries and dissociation energies
 with AE present and those resulting from an  eCEPP representation of core electrons.
}
\end{table*}

The test set used is composed of the molecules and states shown in
Table\ \ref{tab:2}.

\subsection{Atomic states used to provide Kleinman-Bylander projectors \label{sec:kb_proj}}

Atomic states for each atom used to generate the projectors used in the Kleinman-Bylander representation
of the eCEPPs are given in Table\ \ref{tab:3}.
All projectors are evaluated using the PBE functional.

\begin{table*}[t]
\begin{tabular}{l@{\hskip 18pt}l@{\hskip 18pt}l@{\hskip 18pt}l}
  \hline \hline
    \multicolumn{1}{l}{Atom}
  & \multicolumn{1}{l}{$s$} & \multicolumn{1}{l}{$p$} & \multicolumn{1}{l}{$d$} \\ \hline
H   &	$1s$   			 &	$2p$    		 &	$3d$			\\
Li  &	[He]$2s$   		 &	[He]$2p$    		 &	[He]$3d$		\\
Be  &	[He]$2s^2$		 &	[He]$2s2p$    		 &	[He]$2s3d$		\\
B   &	[He]$2s^22p$    	 &	[He]$2s^22p$    	 &	[He]$2s^23d$		\\
C   &	[He]$2s^22p^2$		 &	[He]$2s^22p^2$		 &	[He]$2s^22p3d$		\\
N   &	[He]$2s^22p^3$		 &	[He]$2s^22p^3$		 &	[He]$2s^22p^23d$	\\
O   &	[He]$2s^22p^4$		 &	[He]$2s^22p^4$		 &	[He]$2s^22p^33d$	\\
F   &	[He]$2s^22p^5$		 &	[He]$2s^22p^5$		 &	[He]$2s^22p^43d$	\\ \hline \hline
    \multicolumn{1}{l}{Atom}
  & \multicolumn{1}{l}{$3s,3p,3d,4s$} & \multicolumn{1}{l}{$4p$} & \multicolumn{1}{l}{$4f$} \\ \hline
Sc  &	[Ar]$3d4s^2$	 &	[Ar]$3d4s4p$	&	[Ar]$3d4f$	\\
Ti  &	[Ar]$3d^24s^2$	 &	[Ar]$3d^24s4p$	&	[Ar]$3d^24f$	\\
V   &	[Ar]$3d^34s^2$	 &	[Ar]$3d^34s4p$	&	[Ar]$3d^34f$	\\
Cr  &	[Ar]$3d^54s$	 &	[Ar]$3d^54p$	&	[Ar]$3d^44f$	\\
Mn  &	[Ar]$3d^54s^2$	 &	[Ar]$3d^54s4p$	&	[Ar]$3d^54f$	\\
Fe  &	[Ar]$3d^64s^2$	 &	[Ar]$3d^64s4p$	&	[Ar]$3d^64f$	\\
Cu  &	[Ar]$3d^104s$	 &	[Ar]$3d^{10}4p$	&	[Ar]$3d^94f$	\\ \hline \hline
\end{tabular}
\caption{ \label{tab:3}
 Atomic states used to generate projectors for the Kleinman-Bylander 
 representation of the eCEPP pseudopotentials.
 States are given for the AE atom, so
 $1s$ does not exist for Li--F pseudo-atoms, and
 $1s$, $2s$, $2p$ do not exist for Sc--Fe and Cu pseudo-atoms.
 Transition metal states with a single occupied $4f$ orbital are 
 $+1$ ions.
}
\end{table*}

\subsection{Basis set optimisation for eCEPPs \label{sec:basis_opt}}

For H, and Li--F we use the generation procedure developed for first row
atoms and the TNDF pseudopotentials by Xu \emph{et al.}\cite{Xu_2013}.
First a HF basis is optimised with 8 Gaussians for the $s$-orbital of H,
and 10 Gaussians for the $s$ and $p$-orbitals of Li--F.
All powers are free to vary.
Ground states energies are minimised for H and B--F, while for Li and Be
the $^2S$ and $^2P$ state energies are minimised separately to provide $s$
and $p$-orbitals (the ground state is $^2S$ for both).
The resulting HF orbitals then provide the pV$n$Z-eCEPP basis set
and contractions.

We then introduce the correlation consistent part of the basis to obtain
cc-pV$n$Z-eCEPP.
This achieved by adding $n-1$ Gaussians to the channels present in
pV$n$Z-eCEPP, followed by adding $n-l+l_0$ Gaussians to the remaining
channels when this is greater than zero (with $l_0=0$ and $1$ for H and Li--F,
respectively).

Added Gaussians are even tempered, and are optimised by minimising
CCSD(T) ground state energies of the neutral atoms for Be--F, and
the homogeneous dimer for H and Li (valence correlation is absent for
H and Li atoms).

Finally, the aug-cc-pV$n$Z-eCEPP basis is generated by adding a single
Gaussian to each channel present in cc-pV$n$Z-eCEPP, and optimising its
power by minimising the CCSD(T) ground state energies for the atomic
anion for H, Li, and B--F, and the BeH anion for Be (Be$^-$ is not stable).

For the $3d$ transition metal atoms we apply the generation procedure of
Balabanov \emph{et al.},\cite{basisDK} but with pseudopotentials
representing the core electrons.
First a HF basis is optimised, with $(12,12,10)$ Gaussians for the $(s,p,d)$
orbitals, and with all powers free to vary.

Minimising the state-averaged HF (SA-HF) energy for [Ar]$4s^23d^{m-2}$
states, with $m=1$--$6$ and $9$ for Sc--Fe and Cu, provides the $s$ part of the
basis.
The optimised basis set for $p$-orbitals is obtained by minimising the SA-HF
energy for [Ar]$s^2d^{m-3}p^1$ states, and the $d$-orbital basis given
by minimising the SA-HF energy of [Ar]$s^2d^{m-2}$,
[Ar]$sd^{m-1}$, and [Ar]$d^m$ states.

The correlation consistent basis is then obtained by adding $n-1$
Gaussians to the HF orbitals, followed by adding $n-l+l_0$ Gaussians to
the remaining channels when this is greater than zero (with $l_0=2$).

Contracting orbitals is somewhat more complex than for the first-rows
atoms.
We start with contractions generated at the HF level of theory, given by
the natural orbitals resulting from state-averaged density matrices.
These provide $(2,2,1)$ contractions from the lowest energy $(s,p,d)$ orbitals.
Natural orbitals are then generated from state-averaged density matrices
resulting from CISD calculations.
These natural orbitals are ordered by decreasing occupation number,
the first $(2,2,1)$ natural orbitals are ignored, and the following
$(n-1,n-1,n-1)$ added to the HF contractions to provide the full set of
contractions for each $n$.

Finally, the aug-cc-pV$n$Z-eCEPP basis is obtained by a single even-tempered
extension of the lowest power in each channel.\cite{basisDK}

\subsection{Core radii for eCEPPs \label{sec:core_radii}}
Core radii used to define the eCEPPs are shown in 
Table\ \ref{tab:4}. 

\begin{table}[t]
\begin{tabular}{lllll} \hline \hline
     & \multicolumn{4}{c}{(\AA)}     \\
Atom & $r^c_s$ & $r^c_p$ & $r^c_d$ & $r^c_f$ \\ \hline
H    & 0.50 & 0.50 & 0.80 & -    \\
Li   & 2.19 & 2.37 & 2.37 & -    \\
Be   & 1.88 & 1.96 & 1.96 & -    \\
B    & 1.41 & 1.41 & 1.41 & -    \\
C    & 1.10 & 1.10 & 1.10 & -    \\
N    & 0.94 & 0.88 & 0.84 & -    \\
O    & 0.80 & 0.75 & 0.99 & -    \\
F    & 0.70 & 0.64 & 0.89 & -    \\
Sc   & 0.94 & 0.96 & 0.94 & 1.05 \\
Ti   & 0.89 & 0.89 & 0.89 & 0.99 \\
V    & 0.86 & 0.86 & 0.86 & 0.96 \\
Cr   & 0.81 & 0.81 & 0.81 & 0.90 \\
Mn   & 0.78 & 0.78 & 0.78 & 0.87 \\
Fe   & 0.76 & 0.73 & 0.76 & 0.84 \\
Cu   & 0.67 & 0.67 & 0.67 & 0.73 \\ \hline \hline
\end{tabular}
\caption{ \label{tab:4}
 Core radii used to define the CEPPs and eCEPPs for
 each atom.
}
\end{table}